\documentclass[submission, Proceedings]{SciPost}

\hypersetup{
    colorlinks,
    linkcolor={red!50!black},
    citecolor={blue!50!black},
    urlcolor={blue!80!black}
}

\usepackage[bitstream-charter]{mathdesign}
\DeclareSymbolFont{usualmathcal}{OMS}{cmsy}{m}{n}
\DeclareSymbolFontAlphabet{\mathcal}{usualmathcal}

\begin{document}

\begin{center}{\Large \textbf{
Anisotropic ionization threshold and directional sensitivity in solid state DM detectors
}}\end{center}

\begin{center}
Matti Heikinheimo\textsuperscript{1$\star$},
Sebastian Sassi\textsuperscript{1},
Kimmo Tuominen\textsuperscript{1},
Kai Nordlund\textsuperscript{1} and
Nader Mirabolfathi\textsuperscript{2}
\end{center}

\begin{center}
{\bf 1} Helsinki Institute of Physics and Department of Physics, University of Helsinki,\\
P.O.Box 64, FI-00014 University of Helsinki, Finland
\\
{\bf 2} Department of Physics and Astronomy and the Mitchell Institute for Fundamental Physics and Astronomy,
Texas A\&M University, College Station, TX 77843, USA
\\
* matti.heikinheimo@helsinki.fi
\end{center}

\begin{center}
\today
\end{center}


\definecolor{palegray}{gray}{0.95}
\begin{center}
\colorbox{palegray}{
  \begin{tabular}{rr}
  \begin{minipage}{0.1\textwidth}
    \includegraphics[width=30mm]{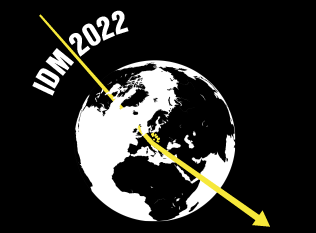}
  \end{minipage}
  &
  \begin{minipage}{0.85\textwidth}
    \begin{center}
    {\it 14th International Conference on Identification of Dark Matter}\\
    {\it Vienna, Austria, 18-22 July 2022} \\
    \doi{10.21468/SciPostPhysProc.?}\\
    \end{center}
  \end{minipage}
\end{tabular}
}
\end{center}

\section*{Abstract}
{\bf
The threshold displacement energy for nuclear recoils depends strongly on the direction of the recoiling nucleus with respect to the crystal lattice. Assuming that similar dependence holds for the ionization threshold for low energy nuclear recoils, we explore the consequences of the resulting directional dependence of the observable event rate in ionization detectors. For low mass dark matter, this effect leads to a daily modulation in the event rate. We discuss how this effect can be utilized to separate the DM signal from the solar neutrino background and how the structure of the modulation signal can be used to identify the type of the DM-nucleon coupling.
}


\section{Introduction}
\label{sec:intro}

To identify the dark matter (DM) signal from the background in direct detection experiments, the time dependence of the signal event rate can be utilized. Annual modulation, resulting from the motion of the earth with respect to the galactic rest frame, is a well known example of this idea. While annual modulation is caused by the changing magnitude of the DM average velocity in the laboratory frame during the year, there is a possibility to observe a daily modulation resulting from the changing direction of this velocity due to the rotation of the earth. This effect arises if the DM scattering rate depends on the direction of the incoming DM particle with respect to the rest frame inherent to the target material. For isotropic targets, such as liquid xenon, no such inherent frame exists and therefore the daily modulation signal is not expected in this case, but solid targets are endowed with a natural frame given by the crystal lattice. For recoil energies not too far above the binding energies of nuclei in their lattice sites the scattering amplitude to a given final state can exhibit directional dependence due to the inherent anisotropy of the crystal structure.

\begin{figure}[tb]
	\begin{center}
		\includegraphics[width=0.37\linewidth]{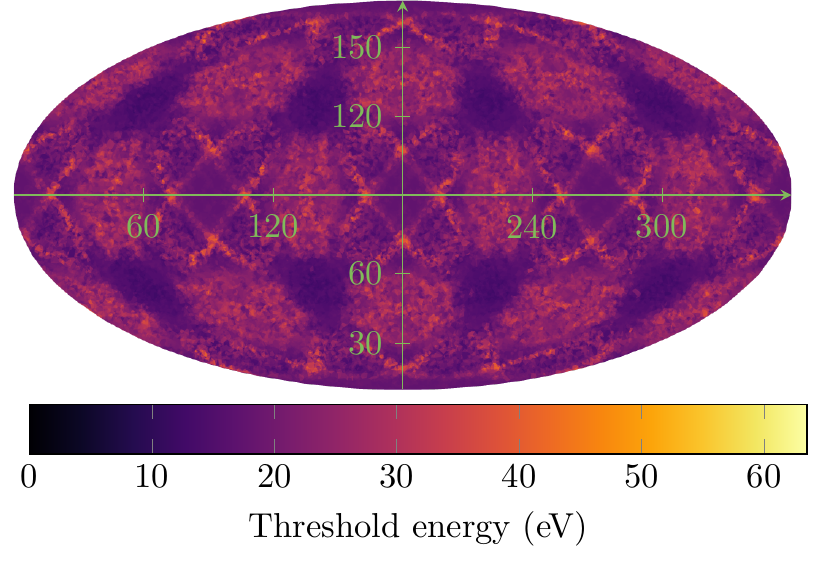}
		\includegraphics[width=0.61\linewidth]{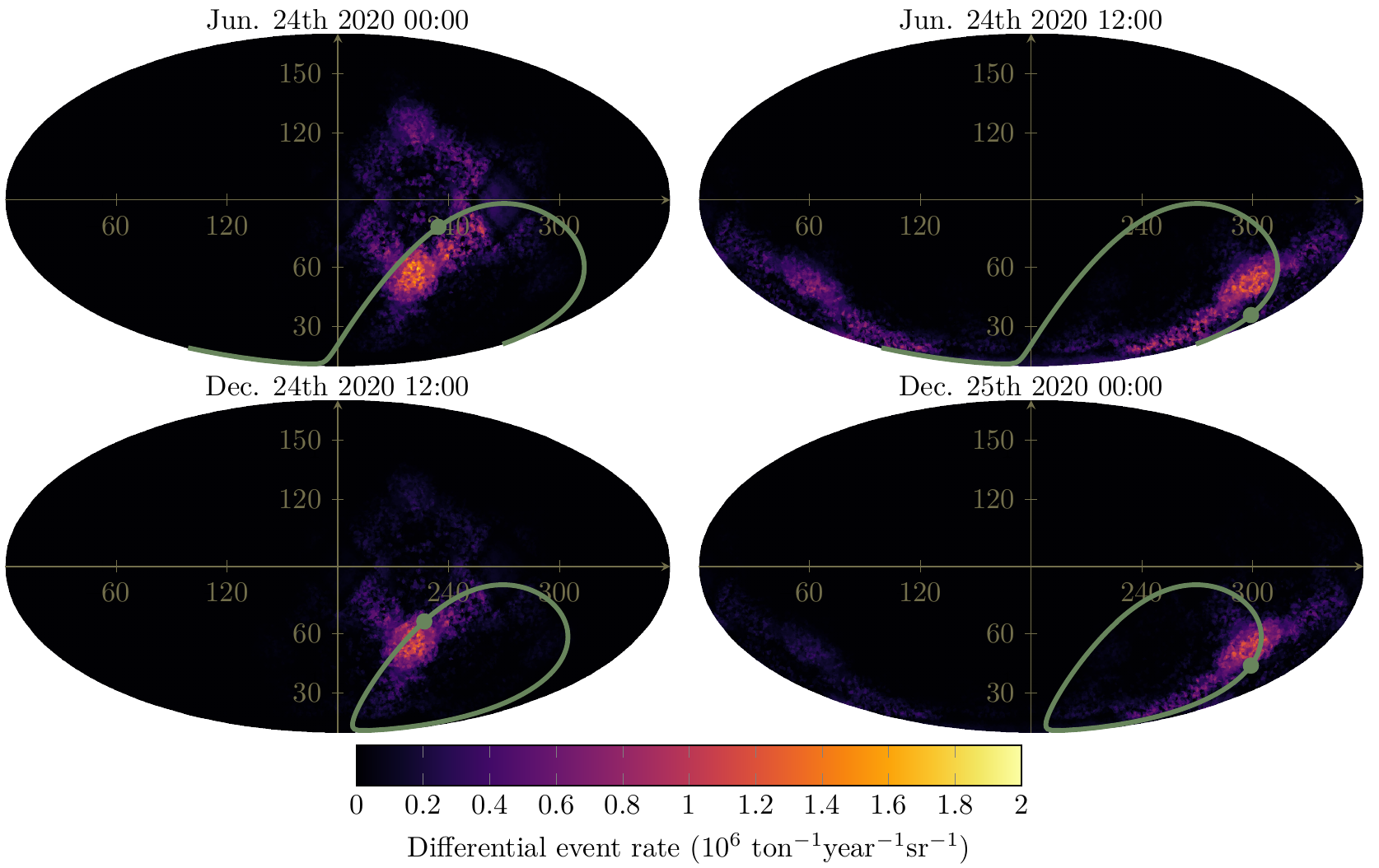}
		\caption{Left: Defect creation threshold energy in a germanium crystal as a function of recoil direction. Right: Differential ionization event rate due to DM recoils at times of daily maxima for a 400 MeV DM particle, assuming spin independent interaction and standard halo model. The figure has appeared before in refrence \cite{Sassi:2021umf}.}
		\label{GeThresholdSurface}
	\end{center}
\end{figure}

A well known example of this phenomenon is the directional dependence of the defect creation threshold in a crystal. The left panel in figure \ref{GeThresholdSurface} shows this threshold energy $E_{\rm thr}(\hat{q})$ in germanium as a function of the recoil direction $\hat{q}$, based on molecular dynamics simulations~\cite{Kadribasic:2017obi,Heikinheimo:2019lwg}. We assume here that the ionization threshold exhibits similar directional dependence. Nuclear recoil events with recoil energy below the ionization threshold will not be observed in an ionization detector. Therefore the observed event rate is given by the integral
\begin{equation}
R=\int\!\!\! d\Omega_q \int_{E_{\rm thr}(\hat{q})}^{E_{\rm{max}}}\!\!\!\!\!\!\!\!\!\!dE\frac{d^2R}{dEd\Omega_q},
\label{eq:directional rate}
\end{equation}
where the lower limit of the energy integral now depends on the recoil direction as shown by the left panel of figure \ref{GeThresholdSurface}. Under standard assumptions for the DM-nucleon interaction and DM velocity distribution, this integral can be performed analytically, as detailed in~\cite{Heikinheimo:2019lwg}. The resulting event rate for a 400 MeV DM particle is shown in the right panel of figure \ref{GeThresholdSurface} at times of daily maxima. The maximum event rate occurs when the direction of the average DM velocity, shown by the green dot, coincides with the low ionization threshold directions in the lattice.

\section{Solar neutrino background}

The ionization event rate due to solar neutrinos also exhibits daily modulation due to the effect described above, as the neutrino flux points away from the Sun, and the inverse direction of the Sun in the lab frame rotates during the day. However, the inverse solar position is always different from the direction of the DM wind, and hence the phase of the modulation signal due to solar neutrinos does not coincide with that of the DM event rate, as described in more detail in~\cite{Sassi:2021umf}. Furthermore, the annual modulation of the solar neutrino signal also differs in magnitude and phase from that of the DM event rate. Therefore the full information contained in the time dependence of the event rate can be used to identify the DM signal even in the presence of the solar neutrino background.

\begin{figure}[tb]
	\begin{center}
		\includegraphics[width=\linewidth]{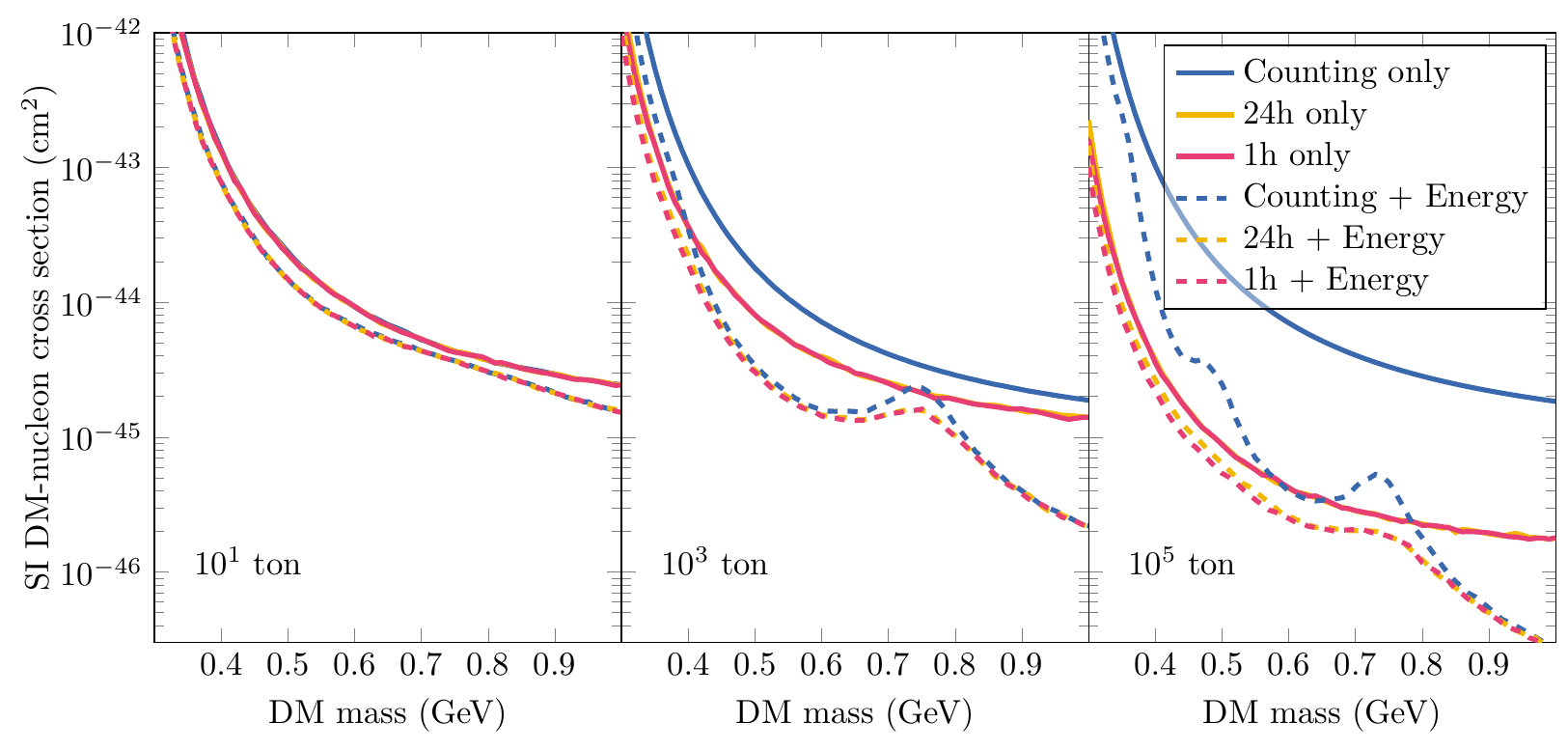}
		\caption{The discovery reach for a germanium detector with the exposure ranging from ten ton years to $10^5$ ton years. The solid lines correspond to experiments using no energy information, whereas dashed lines include full energy information.
The colors correspond to the resolution of time information used from no time information (blue), to 24h binning (yellow), and 1h binning (red). The figure has appeared before in refrence \cite{Sassi:2021umf}.}
		\label{fig:neutrinofloor}
	\end{center}
\end{figure}

The discovery reach of a germanium ionization experiment, assuming no other background beyond the solar neutrinos, is shown in figure \ref{fig:neutrinofloor}, for various assumptions of the exposure, energy resolution, and event time binning. We find that the improvement in the reach due to the daily modulation effect (red lines), compared to the reach obtained with the annual modulation (yellow lines) is modest at best. However, as described in~\cite{Sassi:2021umf}, we find that the daily modulation alone provides similar gain in reach as the annual modulation effect. Utilizing both modulation signals can therefore provide redundancy and a possibility to cross-check the presence of the modulation signal in the event rate. As discussed recently, a spurious annual modulation signal can be induced by time-dependent backgrounds if the averaging procedure does not properly account for the time dependence~\cite{Buttazzo:2020bto,Messina:2020pnt,COSINE-100:2022dvc}. Therefore this redundancy could be very useful for controlling systematics and confirming the galactic origin of the signal.

\section{Identifying the DM-nucleon coupling}

\begin{figure}[tb]
	\begin{center}
		\includegraphics[width=0.4\linewidth]{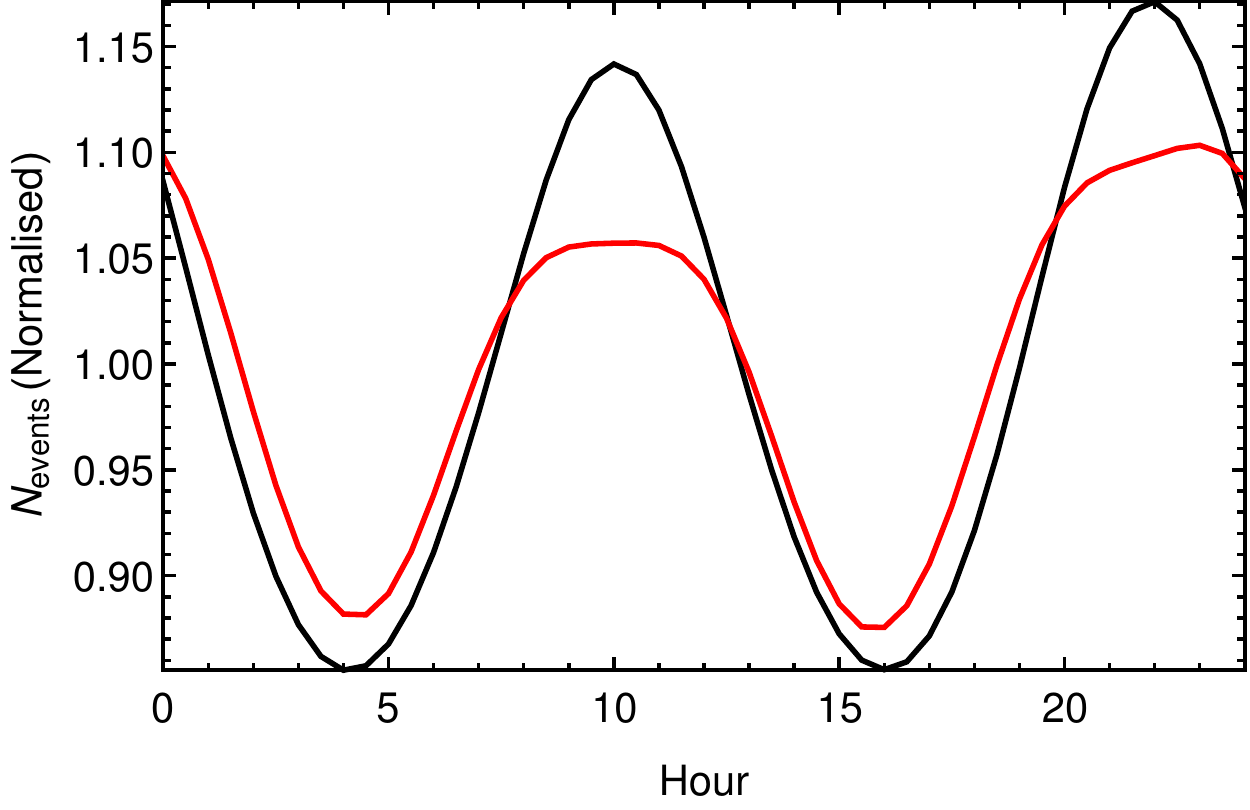}
		\includegraphics[width=0.4\linewidth]{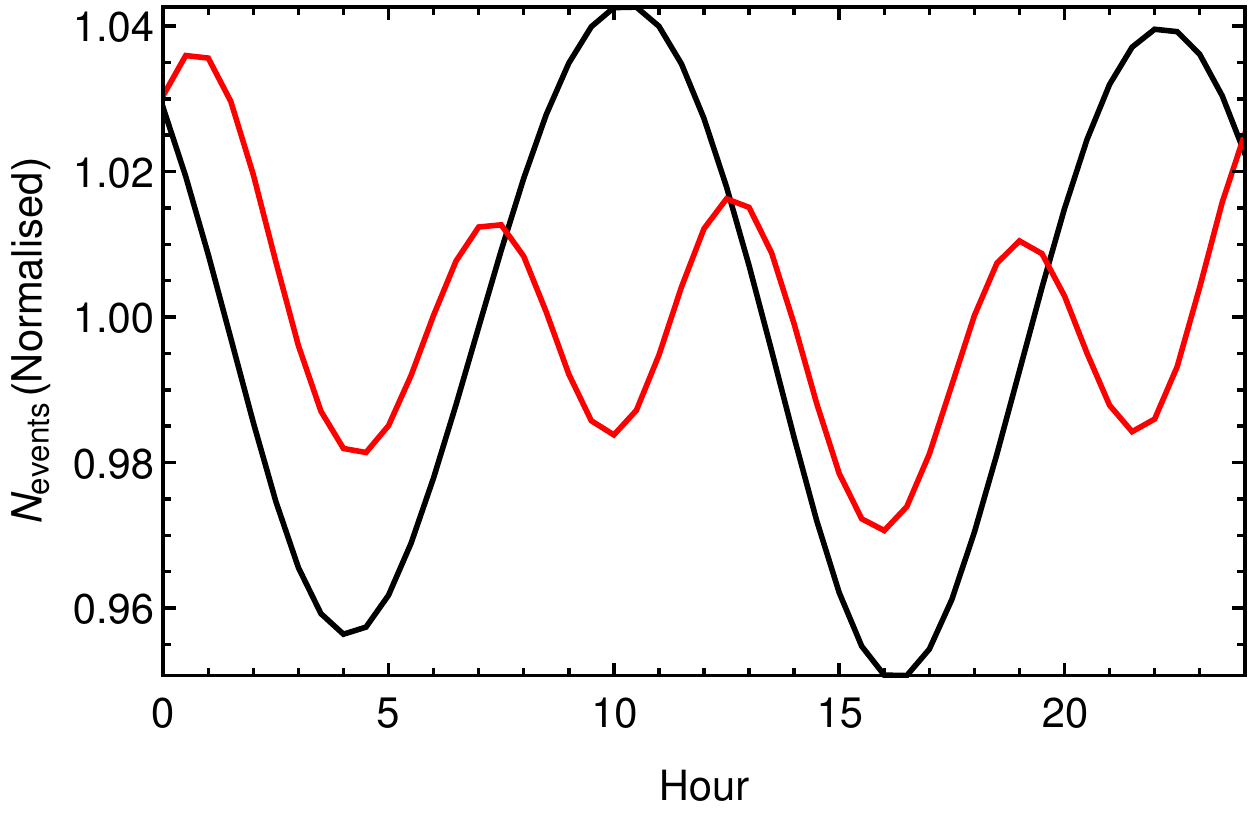}
		\caption{Normalised event rate $R(t)/\langle R \rangle$ for $v^0$ (black) and $v_\perp^2$ (red) interactions as
a function of time for $m_{DM}=0.33$ GeV (left), $m_{DM}=0.36$ GeV (right). Figure adapted from reference \cite{Heikinheimo:2019lwg}.}
		\label{fig:dailymod_transverse}
	\end{center}
\end{figure}

Beyond the simplest spin-independent DM-nucleon coupling assumed above, the coupling can take various forms~\cite{Kavanagh:2015jma,Fitzpatrick:2012ix,DelNobile:2018dfg}, affecting the angular distribution of the recoil rate $dR/d\Omega_q$. As discussed in more detail in~\cite{Heikinheimo:2019lwg}, this in turn affects the structure of the daily modulation signal of the DM event rate in an ionization detector. We demonstrate this effect in figure \ref{fig:dailymod_transverse}, where the time dependent ionization rate in germanium is shown for the standard spin-independent interaction in black, and for a velocity-dependent interaction in red. The left figure shows the rate for a 330 MeV DM particle, and the right figure for 360 MeV mass.

As explained in~\cite{Heikinheimo:2019lwg}, the behaviour seen in the figure is an example of a general trend: Lighter DM mass implies larger modulation amplitude, but less difference in the structure of the modulation signal between different DM-nucleon couplings. For heavier mass the difference in the shape of the modulation signal grows, but the amplitude decreases. Therefore there is a window in DM mass, where the identification of the coupling operator is feasible using the daily modulation data, above which the modulation amplitude vanishes, and below which the information contained in the structure of the signal vanishes, along with overall deacrease in the event rate. However, this window is specific to the detector material. Thus a larger range of masses can be spanned by utilizing detectors with different target materials.

\section{Conclusion}
If the ionization threshold due to nuclear recoils in a semiconductor crystal exhibits directional dependence similar to the defect creation threshold energy, this will result in a directional dependence of the ionization event rate. We have analyzed the daily modulation in the ionization event rate due to DM recoils in a germanium targer resulting from this threshold effect. The daily modulation signal can be used to cross check the galactic origin of an annual modulation signal, providing similar gain in background rejection against solar neutrinos. Furthermore, the structure of the signal depends on the DM-nucleon coupling. Therefore analyzing the shape of the modulation signal can reveal information about the properties of the DM particle beyond the mass and coupling strength, that could be utilized to identify the particle physics description of DM.

\paragraph{Funding information}
This work has been supported by the Academy of Finland project $\# 342777$.



\bibliography{IDMproceedings.bib}
\bibliographystyle{SciPost_bibstyle}

\nolinenumbers

\end{document}